\documentclass{article}
\usepackage{amssymb,latexsym}
\usepackage{amsmath, amsthm}
\newcommand{\be}{\begin{equation}}
\newcommand{\ee}{\end{equation}}

\input xy
\xyoption{all}

\begin{document}

\title{Local approach to thermodynamics of black holes}

\author{\it Ewa Czuchry$^1$\\
\rm $^1$ Yukawa Institute for Theoretical Physics, Kyoto
University, \\ Kitashirakawa-Oiwake-Cho, Sakyo-ku, Kyoto 606-8502,
Japan\\E-mail: eczuchry@yukawa.kyoto-u.ac.jp\\
\it Jacek Jezierski$^2$\thanks{supported by the Polish Research
Council grant KBN 2 P03B 073 24}  \\
\rm $^2$Department of Mathematical Methods in Physics, \\
University of Warsaw,
ul. Ho\.za 69, 00-682 Warszawa, Poland\\
E-mail: Jacek.Jezierski@fuw.edu.pl\\
\it Jerzy Kijowski$^3$\\
 $^3$Center for Theoretical Physics, Polish Academy of Sciences, \\
 Al. Lotnik\'ow 32/46; 02-668
Warszawa, Poland\\
E-mail: kijowski@cft.edu.pl }


\maketitle

\abstract{ Hamiltonian description of gravitational field
contained in a spacetime region with boundary $S$ being a
null-like hypersurface (a wave front) is discussed. Complete
generating formula for the Hamiltonian dynamics (with no surface
integrals neglected) is presented. A quasi-local proof of the 1-st
law of black holes thermodynamics is obtained as a consequence, in
case when $S$ is a non-expanding horizon. The 0-th law and Penrose
inequalities are discussed from this point of view.}



\section{Introduction}

Evolution of gravitational field within a finite tube with a {\em
time-like boundary} was considered in \cite{pieszy} and then
reformulated in \cite{K1}. Here, we extend this description to the
case of a {\em wave front} (a three-dimensional submanifold whose
internal metric is degenerate). Restricting our result to the
special case of wave fronts, namely to non-expanding horizons, we
obtain the 1-st law of thermodynamics of black holes as a simple
consequence.

Hamiltonian formulation of any field theory needs always an
appropriate control of boundary data. To illustrate this point,
consider as an example the linear theory of an elastic string.
Field configuration of a string is described by its displacement
function: ${\mathbb R} \times [a , b ] \ni (t,x) \rightarrow
\varphi(t,x) \in {\mathbb R}\ ,$ fulfilling the wave equation,
where velocity ``$c$'' is a combination of the string's proper
density (per unit length) and its elasticity coefficient:
\begin{equation}\label{wave}
    \frac 1{c^2} \frac {\partial^2}{\partial t^2}\ \varphi =
    \frac {\partial^2}{\partial x^2}\ \varphi \ .
\end{equation}
Passing to appropriate time and length units, we may always put
$c=1$. The dynamics of the system may be derived from the
Lagrangian density
\begin{equation}\label{Lagr}
    L = - \frac 12 \sqrt{|\det g |} \  g^{\mu\nu}
    (\partial_\mu \varphi) (\partial_\nu \varphi)
    =  \frac 12 \sqrt{|\det g |} \  \left(
    (\dot\varphi )^2 - (\varphi^\prime)^2 \right)
    \ ,
\end{equation}
where $\mu,\nu = 0,1$ and $(x^0,x^1)=(t,x)$, $g_{\mu\nu}=$
diag$(-1,+1)$, ``dot'' denotes the time derivative and ``prime''
denotes the space derivative. Entire information about the
dynamics of the string may be encoded in the form of the following
Lagrangian generating formula (see \cite{Kij-Tulcz} or
\cite{springer} for its correct symplectic interpretation):
\begin{equation}\label{gen-Lagr-true}
  \delta L(\varphi , \partial_\nu \varphi ) =
  \partial_\mu \left( p^\mu \delta \varphi \right)=
  \left( \partial_\mu p^\mu \right) \delta \varphi +
  p^\mu \delta \left( \partial_\mu \varphi \right) \ .
\end{equation}
It contains the definition of the canonical momenta: 1) the
kinetic momentum
\[
   \pi := p^0 = \frac {\partial L}{\partial (\partial_0\varphi)}
   =  \partial_0\varphi = \dot\varphi \ ,
\]
and 2) the stress density
\[
   \pi^\perp := p^1 = \frac
   {\partial L}{\partial (\partial_1\varphi)}
   = -\partial_1\varphi = - \varphi^\prime \ ,
\]
together with the Euler-Lagrange equation, obviously equivalent to
(\ref{wave}):
\[
   \partial_\mu p^\mu =
   \frac {\partial L}{\partial \varphi} \ .
\]
Integrating infinitesimal generating formula (\ref{gen-Lagr-true})
over the entire string $[a,b]$ we obtain the finite generating
formula:
\begin{equation}\label{finite-L}
    \delta \int_a^b L = \int_a^b \left( \dot\pi \delta\varphi
    + \pi \delta\dot\varphi \right) +
    \left[
    \pi^\perp \delta\varphi \right]^b_a \ .
\end{equation}
Hamiltonian description of the same dynamics is obtained {\em via}
Legendre transformation between $\pi$ and $\dot\varphi$, putting
$\pi \delta\dot\varphi = \delta (\pi \dot\varphi)- \dot\varphi
\delta \pi$:
\begin{equation}\label{finite}
    -\delta {\cal H} =
    \int_a^b \left( \dot\pi \delta\varphi
    - \dot\varphi \delta \pi \right) +
    \left[
    \pi^\perp \delta\varphi \right]^b_a \ ,
\end{equation}
with
\begin{eqnarray}\label{Hamilt-H}
  {\cal H}:=\int_a^b (\pi\dot\varphi - L)
  = \frac 12 \int_a^b (\pi^2 + (\varphi^\prime)^2)  \ .
\end{eqnarray}
This formal expression acquires a precise, infinitely-dimensional,
Hamiltonian meaning:
\begin{eqnarray}\label{Hamilt-fi}
  {\dot \pi}  =  - \frac {\delta {\cal H}}{\delta\varphi}\ \ \ , \ \
  \ {\dot \varphi}  =  \frac {\delta {\cal H}}{\delta
  \pi}\ \ \ ,
\end{eqnarray}
as soon as the boundary terms in (\ref{finite}) are killed by
imposing e.g.~the Dirichlet boundary conditions, i.e. by
restricting ourselves to an infinitely dimensional phase space of
initial data $( \varphi , \pi )$, defined on $[a,b]$ and
fulfilling conditions: $\varphi (a) \equiv A$, $\varphi (b) \equiv
B$. Within this phase space we have
$\delta\varphi(a)=\delta\varphi(a)=0$ and equations
(\ref{Hamilt-fi}) hold.

Consider now the subspace of static solutions: ${\dot \pi} = 0 =
{\dot \varphi}$. Due to (\ref{Hamilt-fi}), these are points where
the derivative of the functional $\cal H$ vanishes and the
Hamiltonian formula (\ref{finite}) reduces to the formula for
virtual work
\begin{equation}\label{virt-work}
    \delta {\cal H} =
    - \left[
    \pi^\perp \delta\varphi \right]^b_a \ ,
\end{equation}
But, due to (\ref{Hamilt-H}), $\cal H$ is manifestly convex. This
implies that every static solution gives the minimal value of the
Hamiltonian in the corresponding phase space. Due to equation
(\ref{wave}) and to boundary conditions, such a solution is given
by: $\pi \equiv 0$ and $\varphi(x)= A + (x-a)\tfrac {B-A}{b-a}$.
Inserting this value into (\ref{Hamilt-H}) we obtain the following
``Penrose-like inequality'':
\begin{equation}\label{ineq}
    \frac {(B-A)^2}{b-a} \le {\cal H} \ ,
\end{equation}
analogous to the gravitational Penrose inequality relating the
energy carried by Cauchy data outside of a horizon $S$ and the
energy of a black hole corresponding to the same value of
appropriate boundary data on $S$.

Instead of controlling the string configuration at the boundary,
we may control e.g.~its stress by applying an appropriate force
$F$. This leads to the Neumann control mode $\pi^\perp(a) = F_{\rm
left}$, $\pi^\perp(b) = F_{\rm right}$, which is again a legitime
Hamiltonian system
\begin{equation}\label{finite-Neu}
    -\delta {\widetilde{\cal H}} =
    \int_a^b \left( \dot\pi \ \delta\varphi
    - \dot\varphi \ \delta \pi \right) -
    \left[
    \varphi \ \delta \pi^\perp \right]^b_a \ ,
\end{equation}
where ${\widetilde{\cal H}}$, obtained {\em via} Legendre
transformation between $\varphi$ and $\pi^\perp$ at the boundary,
plays role of a free energy:
\begin{eqnarray}\label{Hamilt-H-tilde}
  {\widetilde{\cal H}}:={\cal H} + \left[
    \varphi \  \pi^\perp \right]^b_a =
    {\cal H} - \left[
    \varphi \  \varphi^\prime \right]^b_a
  = \frac 12 \int_a^b (\pi^2 - (\varphi^\prime)^2
  - 2 \varphi\varphi^{\prime\prime})  \ .
\end{eqnarray}
Again, the boundary term in (\ref{finite-Neu}) vanishes due to
Neumann conditions and the field dynamics reduces to
(\ref{Hamilt-fi}). However, the new Hamiltonian
(\ref{Hamilt-H-tilde}) is obviously non-convex. There is,
therefore, no ``Penrose-like'' inequality in this mode: static
solutions corresponding to stationary points of the Hamiltonian
${\widetilde{\cal H}}$ {\em are not} minimal points of such a
``free energy''.

In \cite{pieszy} the dynamics of the gravitational field within a
time-like world tube $S$ was analyzed in a similar way. For this
purpose the so called ``affine variational principle'' was used,
where the Lagrangian function depends on the Ricci tensor of a
spacetime connection $\Gamma$. In this picture, the metric tensor
$g$ arises only in the Hamiltonian formulation as the momentum
canonically conjugate to $\Gamma$. Later, it was proved in
\cite{K1} that the Hamiltonian dynamics obtained this way is
universal and does not depend upon a specific variational
formulation we start with (actually, it can be derived from field
equations only, without any use of variational principles, the
existence of them being a consequence of the ``reciprocity'' of
Einstein equations -- see \cite{Kij-Tulcz} and \cite{springer}).
On the contrary, the Hamiltonian picture is very sensitive to the
method of controlling the boundary data. A list of natural control
modes, leading to different ``quasilocal Hamiltonians'', is given
in \cite{K1}. A conjecture about the ``true mass'', based on an
analysis of the linearized theory \cite{JJGRG31}, is also
formulated there.

The aim of the present paper is to give a generalization of the
above results to the case when the boundary $S$ is a
three-dimensional submanifold of spacetime $M$ whose internal
three-metric $g_{ab}$ is degenerate.

\section{Dynamics of the gravitational field inside a null
hypersurface}

Consider gravitational field dynamics inside a null hypersurface
$S$:

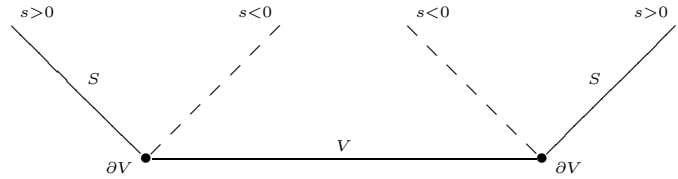
\[
\xymatrix@M=0pt@R=1.7cm@C=1.7cm{ \ar@{-}[dr]_>{\partial V
}^<{s>0}^S & &\ar@{--}[dl]_<{s<0}&\ar@{--}[dr]^<{s<0}&&
\ar@{-}[dl]^>{\partial V}_<{s>0}_S\\
&{\bullet}\ar@{-}[r]&\ar@{-}[r]^V&\ar@{-}[r]&{\bullet}&}\]

\vskip 0.5cm

Parameter $s = \pm 1$ labels two possible situations: an expanding
or a shrinking wave front (if $S$ is a horizon, these correspond
to a black hole or a white hole case). To simplify notation we use
coordinates $x^\mu$, $\mu=0,1,2,3$, adapted to the above
situation: $x^0 = t$ is constant on a chosen family of Cauchy
surfaces whereas $x^3$ is constant on the boundary $S$ (this does
not mean that $x^3$ is null-like everywhere, but only on $S$).
Coordinates $x^A$, $A=1,2$, are ``angular'' coordinates on the
2-surface $\partial V =V\cap S$ whose topology is assumed to be
that of a 2-sphere. Finally, $x^k$, $k=1,2,3$, are spatial
coordinates on the Cauchy surfaces $\{ x^0 =$ const.$\}$ and
$x^a$, $a=0,1,2$, are coordinates on the boundary $S$.

In paper \cite{Bl-H} we derive the following formula:
\begin{align}
- \delta {\cal H} & =  \frac 1{16 \pi} \int_V  \left( {\dot
P}^{kl}  \delta g_{kl} - {\dot g}_{kl} \delta P^{kl} \right) +
\frac s{8 \pi} \int_{\partial V} ( {\dot \lambda} \delta  a  -
{\dot  a} \delta \lambda ) \nonumber \\
 & +  \frac s{16 \pi}
\int_{\partial V} \left( \lambda {l}^{AB}\delta g_{AB}
 -2\left(w_0\delta\Lambda^0-\Lambda^A\delta w_A
\right)\right)\ , \label{form-zerowa1a}
\end{align}
where
\begin{equation}
{\cal H}  =   \frac 1{8 \pi} \int_V {\cal G}^0_{\ 0} + \frac s{8
\pi}
 \int_{\partial V}  \lambda l
\equiv   \frac s{8 \pi}
 \int_{\partial V}  \lambda l
  \, , \label{H-grav-volume1}
\end{equation}
and $P^{kl}$ denotes external curvature of the Cauchy surface,
written in ADM form. Moreover, $\lambda=\sqrt{\det g_{AB}}$ is a
two-dimensional volume form and $a=-\frac12\log |g^{00}|$. The
remaining objects are constructed from a null field $K$ tangent to
$S$. It is not unique, since $fK$ is also a null field for any
function $f$ on $S$. For purposes of the Hamiltonian formula
(\ref{form-zerowa1a}) we always use the normalization compatible
with the (3+1)-decomposition used here: $<K,dx^0>\; =1$. Hence,
$K=\partial_0 - n^A \partial_A$. The vector-density $\Lambda^a =
\lambda K^a =(\lambda,-\lambda n^A)$ is uniquely defined on $S$.
Now we define\footnote{This is the (2+1)-decomposition of the
extrinsic curvature $Q^a{_b}(K)$ defined in \cite{jjrw}.}
\begin{equation}\label{el-def}
l_{ab}:=  -g(\partial_b,\nabla_a K) = - \frac 12 \pounds_K g_{ab}
\ ,
\end{equation}
\begin{equation}\label{wu-def}
w_a := -<\nabla_a K , dx^0> \ ,
\end{equation}
where $g_{ab}$ is the induced (degenerate) metric on $S$. Denoting
by $\tilde{\tilde{g}}^{AB}$ the inverse two-metric, we define the
null mean curvature: $l=\tilde{\tilde{g}}^{AB}l_{AB}$ (often
denoted by $\theta$ -- see \cite{jjrw}).

The volume term in (\ref{H-grav-volume1}) vanishes due to
constraint equations  ${\cal G}^0_{\ \nu}=0$\footnote{In the
presence of matter the volume term equals ${\cal G}^0_{\ 0}-8\pi
T^0_{\ 0}$ and also vanishes due to constraint equations.}. ${\cal
G}^0_{\ 0}$ is often denoted by $N{ H}+N^k{ H}_k$ (see
e.g.~\cite{Misner}), where $ H$ is the scalar (``Hamiltonian'')
constraint and ${ H}_k$ are the vector (``momentum'') constraints,
$N$ and $N^k$ are the lapse and the shift functions. Constraint
equations ${ H}=0$ i ${ H}_k=0$ imply vanishing of ${\cal G}^0_{\
0}$.

In \cite{Bl-H} we give two independent proofs of the formula
(\ref{form-zerowa1a}). The first one is analogous to the
transition from formula (\ref{gen-Lagr-true}) to formula
(\ref{finite}). For this purpose we use Einstein equations written
analogously to (\ref{gen-Lagr-true}) (cf.~\cite{pieszy}):
\begin{equation}\label{gen-Lang-grav}
  \delta L = \partial_\kappa \left(
  \pi^{\mu\nu}\delta
   A^\kappa_{\mu\nu} \right)
 \ ,
\end{equation}
where $\displaystyle {\pi}^{\mu\nu} := \frac 1{16 \pi} \sqrt{|g|}
\ g^{\mu\nu}$, and $\displaystyle A^{\lambda}_{\mu\nu} :=
{\Gamma}^{\lambda}_{\mu\nu} - {\delta}^{\lambda}_{(\mu}
{\Gamma}^{\kappa}_{\nu ) \kappa}$.

Integrating (\ref{gen-Lang-grav}) over a volume  $V$ and using
metric constraints for the connection $\Gamma$, we directly prove
(\ref{form-zerowa1a}). However, an indirect proof is also
provided, based on a limiting procedure, when a family
$S_\epsilon$ of time-like surfaces tends to a light-like surface
$S$. It is shown that the non-degenerate formula derived in
\cite{pieszy} and \cite{K1} gives (\ref{form-zerowa1a}) as a
limiting case for $\epsilon \rightarrow 0$.

The last term in (\ref{form-zerowa1a}) may be written in the
following way
\[
-\Lambda^A\delta w_A =\lambda n^A \delta w_A =   n^A \delta {\cal
W}_A - n^A w_A \delta \lambda \ ,
\]
where ${\cal W}_A := \lambda w_A$ and
$n^A:=\tilde{\tilde{g}}^{AB}g_{0B}$.
Denoting  $\kappa :=  n^A w_A -w_0  = - K^a w_a$ we finally obtain
the following generating formula:
\begin{align}
   - \delta {\cal H} & =  \frac 1{16 \pi} \int_V  \left( {\dot
   P}^{kl}  \delta g_{kl} - {\dot g}_{kl} \delta P^{kl} \right) +
   \frac s{8 \pi} \int_{\partial V} ( {\dot \lambda} \delta  a  -
   {\dot  a} \delta \lambda )
   \label{17}
   \\
   & +  \frac s{16 \pi}
  \int_{\partial V} \left( \lambda {l}^{AB}\delta g_{AB}
  +2\left(\kappa\delta\lambda - n^A\delta {\cal
  W}_A \right)\right)\ . \label{form-zerowa1}
\end{align}
Quantity $\kappa$ fulfilling: $K^a\nabla_a K= \kappa K$,  is
traditionally called a ``surface gravity'' on $S$. Its value is
not an intrinsic property of the surface itself, but depends upon
a choice of the null field  $K$ on $S$ (i.e. the
(3+1)-decomposition of spacetime). In black hole thermodynamics
there is a privileged time, compatible with the Killing field of
stationary solution and normalized to unity at infinity. In this
case the above formula provides, as will be seen, the so called
first law of black hole thermodynamics.

We stress that the symplectic structure of Cauchy data, given by
two integrals in (\ref{17}), is invariant with respect to
spacetime diffeomorphisms (see \cite{K1}). Neglecting the last,
surface integral and defining symplectic form only by the volume
integral destroys this gauge invariance.

\section{Dynamics of gravitational field outside the null surface}

Consider now dynamics of the gravitational field outside a wave
front $S^-$. We first add an external, timelike (non-degenerate)
boundary $S^+$ and the situation is illustrated by the following
figure:
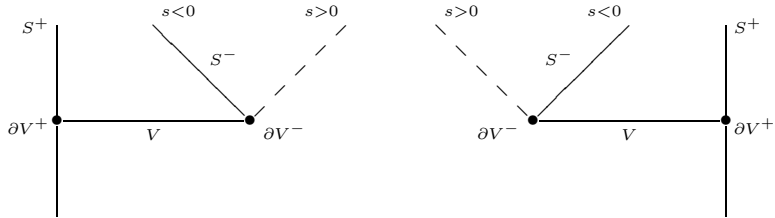
\[ \hspace*{-0.2cm}
\xymatrix@M=0pt@R=1.2cm@C=1.2cm{ \ar@{-}[d]_<{S^+} &
\ar@{-}[dr]^<{s<0}^{S^-}& & \ar@{--}[dl]^>{\partial V^-
}_<{s>0}&\ar@{--}[dr]_>{\partial V^- }^<{s>0}&&
\ar@{-}[dl]_<{s<0}_{S^-} &\ar@{-}[d]^<{S^+}\\
{\bullet}\ar@{-}[r]_>V& \ar@{-}[r]&{\bullet}&&&{\bullet}\ar@{-}[r]
& \ar@{-}[r]_<V&{\bullet}\\ \ar@{-}[u]^>{\partial
V^+}&&&&&&&\ar@{-}[u]_>{\partial V^+} }\]
where $\partial V^+= V\cap S^+$, and $\partial V^-=V\cap S^-$.
Because $\partial V^- $ enters with negative orientation, we have:
$\int_{\partial V}=\int_{\partial V^+}-\int_{\partial V^- }$.
Integrating again Einstein equations written in the form
(\ref{gen-Lang-grav}) over $V$, using techniques derived in
\cite{pieszy} and \cite{K1} to handle surface integrals over
$\partial V^+$ and formula (\ref{form-zerowa1a}) to handle the
surface integrals over $\partial V^-$, we obtain:
\begin{align}\label{form-zerowa2}
- \delta {\cal H} & = -\delta {\cal H^+}-\delta {\cal H^-} = \frac
1{16 \pi} \int_V \left( {\dot P}^{kl}  \delta g_{kl} - {\dot
g}_{kl} \delta P^{kl} \right) \nonumber \\
 &\hspace{-0.5cm} + \frac{1}{8\pi}\int_{\partial V^+
}\left( \dot{\lambda}\delta\alpha
  -\dot{\alpha}\delta\lambda\right)
+ \frac s{8 \pi} \int_{\partial V^- } ( {\dot \lambda} \delta a  -
{\dot  a} \delta \lambda )
 -\frac{1}{16\pi}\int_{\partial V^+ } \mathcal{Q}^{ab}\delta g_{ab}
 \nonumber \\
 &
+  \frac s{16 \pi} \int_{\partial V^- } \left( \lambda
{l}^{AB}\delta g_{AB}
 -2\left(w_0\delta\Lambda^0-\Lambda^A\delta w_A
\right)\right)\ ,
\end{align}
where $\alpha$ is the ``hyperbolic angle'' between $V$ and $S^+$,
whereas $\mathcal{Q}^{ab}$ is the external curvature of $S^+$
written in the ADM form (cf. \cite{K1}). The contribution ${\cal
H^+}$ to the total Hamiltonian from the external boundary is
written here in the form of a ``free energy'' proposed in
\cite{K1}:
\begin{equation}
{\cal H^+}  = -\frac{1}{8\pi}\int_{\partial V^+} {\mathcal{Q}^0}_0
- E_0 \ ,
\end{equation}
where the additive gauge $E_0$ is chosen in such a way that the
entire quantity vanishes if $\partial V^+$ is a round sphere in a
flat space. The internal contribution to the energy is given by
formula (\ref{H-grav-volume1}) with $\partial V$ replaced by
$\partial V^-$. Shifting the external boundary to space infinity:
$\partial V^+ \rightarrow \infty$, the external energy ${\cal
H}^+$ gives the ADM mass, which we denote by $\cal M$, whereas the
remaining surface integrals over $\partial V^+$ vanish. This way
we obtain the following generating formula for the field dynamics
outside of an arbitrary wave front $S^-$ in an asymptotically flat
spacetime:
\begin{align}\label{form-zerowa21}
 -\delta {\cal M} -\delta{\cal H^-} &= \frac 1{16 \pi} \int_V
\left( {\dot P}^{kl}  \delta g_{kl} - {\dot g}_{kl} \delta P^{kl}
\right) + \frac s{8 \pi} \int_{\partial V^- } ( {\dot \lambda}
\delta a - {\dot  a} \delta \lambda )\nonumber \\ & +  \frac s{16
\pi} \int_{\partial V^- } \left( \lambda {l}^{AB}\delta g_{AB}
 +2\left(\kappa\delta\lambda - n^A\delta {\cal
W}_A \right)\right)\ .
\end{align}

\section{Black hole thermodynamics}\label{czdz}
In this Section we apply the above result to the situation, when
the wave front $S^-$ is a non-expanding horizon, i.e. $l=0$ (see
\cite{jjrw}). In this case the ``internal energy''  ${\cal H^-}$
vanishes. Moreover, Einstein equations imply ${l}^{AB}=0$ (see
\cite{JKC}) and the definition of $w_a$ reduces to: $\nabla_a K =
-w_a K$. We obtain the following generating formula for the black
hole dynamics
\begin{align}
 - \delta {\cal M} & =  \frac 1{16 \pi} \int_V \left( {\dot P}^{kl}
 \delta g_{kl} - {\dot g}_{kl} \delta P^{kl} \right)  + \frac s{8
 \pi} \int_{\partial V^-} ( {\dot \lambda} \delta a  - {\dot  a}
 \delta \lambda )
 \nonumber \\
 &  +  \frac s{8\pi} \int_{\partial V^-}
 \left(\kappa\delta\lambda - n^A\delta
  {\cal W}_A \right)\ , \label{form-zerowa3}
\end{align}
where $s=1$ for a white hole, and $s=-1$ for a black hole.

The so called "black hole thermodynamics" consists in analysing
possible stationary situations. By stationarity we understand the
existence of a timelike symmetry (Killing) vector field. If such a
field exists,  we may always choose a coordinate system such that
the Killing field becomes $\tfrac{\partial}{\partial x^0}$ and all
the time derivatives (dots) vanish. Hence, formula
(\ref{form-zerowa3}) reduces to:
\begin{equation}\label{form-zerowa11}
 \delta {\cal M}  =
   - \frac s{8 \pi}
  \int_{\partial V^-} \left(\kappa\delta\lambda - n^A\delta
  {\cal W}_A \right)\ .
\end{equation}
We assumed here that $\tfrac{\partial}{\partial x^0}$ is tangent
to $S$. If this is not the case, we would have a one-parameter
family of horizons. Such phenomenon corresponds to the Kundt's
class of metrics (see e.g.~\cite{TPJLJJ}). The known metrics of
this class are not asymptotically flat. We do not know whether or
not this is a universal property and we exclude such a pathology
by the above assumption.

We have shown in \cite{JKC} that there is a canonical affine
fibration $\pi : S \rightarrow B$ over a base manifold $B$, whose
topology is assumed to be that of a sphere $S^2$. The affine
structure of the fibers is implied by the fact that they are
null-geodesic lines in $M$. Identity $-2{l}_{ab}={\pounds}_K
g_{ab}=0$ implies that the metrics $g$ on $S$ may be projected
onto the base manifold $B$, which acquires a Riemannian two-metric
tensor $h_{AB}$. The degenerate metric $g_{ab}$ on  a manifold $S$
is simply the pull back of $h_{AB}$ from $B$ to $S$: $g = \pi^* h
\ .$

The quantity $w_a$  is not an intrinsic property of the surface
itself, but depends upon a choice of the null field  $K$ on $S$.
Indeed, if ${\tilde K} = \exp (-\gamma)K$ then $\tilde{w}_a= w_a+
\partial_a \gamma$. In particular, there are on $S$ vector fields
$K$ such that $K^a \nabla_a K = 0$ and, consequently, $\kappa=0$.
These are null geodesic fields tangent to fibers of $\pi : S
\rightarrow B$.

In case of a black hole, there is a privileged field $K$,
compatible with the time-like symmetry of the solution, which is
normalized to unity at infinity. This way the quantities $\kappa$
and $w_A$ in formula (\ref{form-zerowa11}) become uniquely
defined.

We have, therefore, two symmetry fields of the metric $g_{ab}$ on
$S$: $\partial_0$ and $K$. Due to normalization chosen above, we
have $<\partial_0 - K , dx^0 > = 0$. Hence, the field
$\vec{n}:=\partial_0 - K = n^A \partial_A$ is purely space-like
and projects on $B$. Moreover, it is a symmetry field of the
Riemannian two-metric $h_{AB}$.

Because the conformal structure of $h_{AB}$ is always isomorphic
to the conformal structure of the unit sphere  $S^2$, we are free
to choose a coordinate system in which  $h_{AB} = f\breve{h}_{AB}$
(and $\breve{h}_{AB}$ denotes the standard unit 2-sphere metrics).
The field $\vec{n}$ is, therefore, the symmetry field of this
conformal structure. Consequently, $\vec{n}$ belongs to the
six-dimensional space of conformal fields on the 2-sphere. Using
remaining gauge freedom, we may choose angular coordinates
$(x^A)=(\theta,\phi)$ in such a way that $\vec{n}$ becomes a
rotation field on the 2-sphere. This means (cf. \cite{lewandowski}
or \cite{Bl-H}) that there exists a coordinate system in which the
following holds:
\begin{equation}\label{Omega1a}
 \vec{n} = -\Omega^k \epsilon_{klm}  {y^l}\partial^m \ .
\end{equation}
Here, $\Omega^k$ are components of a three-dimensional vector
called  angular velocity of the black hole, and  $y^k$ are
functions on $S^2$ created by restricting Cartesian coordinates on
$\mathbb{R}^3$ to a unit 2-sphere. We can also set $z$-coordinate
axis parallelly to angular velocity vector field. After a suitable
rotation we have: $(\Omega^k)=( 0,0,\Omega)$, $z=y^3=\cos\theta$,
and:
\begin{equation}\label{Omega1}
  \vec{n}= - \Omega\frac{\partial}{\partial\varphi}\ .
\end{equation}
Inserting this into (\ref{form-zerowa11}) we obtain
\begin{equation}\label{calka}
  -\frac1{8\pi}\int_{\partial V^-} n^A\delta
  {\cal W}_A =  \Omega \delta J \ ,
\end{equation}
where
\begin{eqnarray}\label{moment-pedu}
  J \equiv J_z: =
  \frac1{8\pi} \int_{\partial V^-} {\cal W}_\varphi  \ ,
\end{eqnarray}
is the $z$-component of the black hole angular momentum.

Up to now we have used only the symmetry of conformal structure
carried by  $h_{AB}$. The symmetry of the metric itself implies
that the conformal factor $f$ is constant along  the field
$\vec{n}$. This follows from the observation that the trace of the
Killing equation implies vanishing of divergence of the field
$\vec{n}$:
\begin{eqnarray}\label{divergencja}
  0 &=& \partial_A (\sqrt{\det h_{CD}}\  n^A)  =
   n^A\sqrt{\det \breve{h}_{CD}}\ \partial_A f
   \ ,
\end{eqnarray}
where the fact that $\vec{n}$ is the symmetry field of the metric
$\breve{h}$ has been used. Formula (\ref{Omega1}) implies that
$\partial_\varphi f =0$ and the conformal factor $f$ must be a
function of the variable $\theta$ only.

It turns out that also its canonical conjugate $\kappa$ may be
gauged in such a way that it is constant along the field $\vec{n}$
(see \cite{Bl-H} for a proof).\footnote{In case $\Omega=0$,
quantities $\kappa$ and $f$ are arbitrary functions on $S^2$.}

This result was obtained locally, or rather   {\em quasi}-locally
-- i.e. from the analysis of the field on the horizon itself.
However, the {\em global} theorems on the existence of stationary
solutions possessing a horizon, imply the so called 0-th law of
thermodynamics of black holes (see \cite{BH}), according to which
{\em the surface gravity $\kappa$ must be constant along the
horizon}. But $A:=\int_{S^2} \lambda$ is the area of the horizon
$S$. Taking this into account and using (\ref{calka}), we derive
from (\ref{form-zerowa11}) the ``first law of black holes
thermodynamics'':
\begin{equation}\label{form-zerowa22}
 -s\delta {\cal M}  =
   \frac 1{8 \pi}{\kappa}\delta A+\Omega\delta J \ .
\end{equation}
Contrary to the theory proposed by Wald and Iyer in \cite{Wald},
the 1-st law (\ref{form-zerowa22}) is, in our approach, a simple
consequence of the complete Hamiltonian formula
(\ref{form-zerowa3}), restricted to the stationary case. As
illustrated by an example of the string dynamics, where formula
(\ref{virt-work}) for virtual work was a consequence of the
Hamiltonian formula (\ref{finite}), a similar ``thermodynamics of
boundary data'' may be expected in any Hamiltonian field theory
(see e.g.~\cite{K1} for the corresponding analysis of the Maxwell
electrodynamics). Also a ``Penrose-like'' inequality (analogous to
(\ref{ineq}) in the string theory) is satisfied as soon as the
Hamiltonian is convex. We very much hope that the gravitational
Penrose inequality can be proved along these lines. Preliminary
results in this direction, based on the analysis of the field
Hamiltonian in linearized gravity (see \cite{JJGRG31}), are
promising.


\end{document}